\documentclass[a4paper]{jpconf}
\usepackage{graphicx}
\begin{document}
\title{Lanczos steps to improve variational wave functions}

\author{Federico Becca$^{1}$, Wen-Jun Hu$^{2}$, Yasir Iqbal$^{3}$, Alberto Parola$^{4}$, Didier Poilblanc$^{5}$, and Sandro Sorella$^{1}$}

\address{$^{1}$ Democritos National Simulation Center, Istituto Officina dei Materiali del CNR and
SISSA-International School for Advanced Studies, Via Bonomea 265, I-34136 Trieste, Italy}
\address{$^{2}$ Department of Physics and Astronomy, California State University, Northridge, 
California 91330, USA}
\address{$^{3}$ Institute for Theoretical Physics and Astrophysics, Julius-Maximilian's University 
of W\"urzburg, Am Hubland, D-97074, W\"urzburg, Germany}
\address{$^{4}$ Dipartimento di Scienza e Alta Tecnologia, Universit\`a dell'Insubria,
Via Valleggio 11, I-22100 Como, Italy}
\address{$^{5}$ Laboratoire de Physique Th\'eorique UMR-5152, CNRS and Universit\'e de Toulouse, 
F-31062 Toulouse, France}

\ead{becca@sissa.it}

\begin{abstract}
Gutzwiller-projected fermionic states can be efficiently implemented within quantum Monte Carlo calculations to 
define extremely accurate variational wave functions for Heisenberg models on frustrated two-dimensional lattices,
not only for the ground state but also for low-energy excitations. The application of few Lanczos steps on top of 
these states further improves their accuracy, allowing calculations on large clusters. In addition, by computing 
both the energy and its variance, it is possible to obtain reliable estimations of exact results. Here, we report 
the cases of the frustrated Heisenberg models on square and Kagome lattices.
\end{abstract}

\section{Introduction}

Obtaining accurate results in strongly-correlated systems remains an extremely difficult task in systems where the 
sign-problem prevents one to use numerically exact quantum Monte Carlo methods.~\cite{Kaul-2013} In the recent past, 
a large variety of numerical approaches have been proposed and devised to study simple models of interacting electrons 
on the lattice. Among the most promising ones, we mention post density-matrix renormalization group (DMRG) techniques, 
based upon tensor networks.~\cite{Schollwoek-2011,Orus-2014} As far as insulating phases are concerned, one important 
issue is to describe Mott insulators, in which the insulating character is driven by the strong electron-electron 
interaction. When lowering the temperature, the general expectation is that some symmetry-breaking phenomena take place,
leading for example to magnetic or Peierls (i.e., dimerized) phases. In two spatial dimensions, a continuous symmetry 
cannot be spontaneously broken at finite temperatures; still, the ground state may possess magnetic long-range order. 
In this regard, the existence of quantum spin liquids, i.e., states that do not break any symmetry down to zero 
temperature, in two dimensions represents a fascinating problem. The key feature to impede long-range ordering is given 
by the presence of frustrating interaction, namely the presence of competing super-exchange couplings that strongly 
enhance quantum fluctuations. The simplest example that captures the low-energy physics of Mott insulators and may give 
rise to spin-liquid ground states is the frustrated Heisenberg model:
\begin{equation}
\label{eqn:heis-ham}
{\cal H} = J_1 \sum_{\langle ij \rangle} {\bf S}_i \cdot {\bf S}_j +
                 J_2 \sum_{\langle\langle ij \rangle\rangle} {\bf S}_i \cdot {\bf S}_j,
\end{equation}
where both $J_1$ and $J_2>0$ and ${\bf S}_i$ are spin-$1/2$ operators at each lattice site $i$; $\langle ij \rangle$ 
and $\langle\langle ij \rangle\rangle$ denote sum over nearest-neighbor and next-nearest-neighbor sites, respectively. 
Here, we consider both the square and Kagome lattices with $N$ sites: $N=L \times L$ for the square lattice and 
$N=3 \times L \times L$ for the Kagome lattice. All energies are given in units of $J_1$.

The ground state properties of these two models have been widely debated in the past twenty years, with contradicting
results. Recent DMRG calculations suggest that the ground state on the Kagome lattice for $J_2=0$ is a gapped $Z_2$ 
spin liquid,~\cite{Yan-2011} while the ground state of the $J_1{-}J_2$ model on the square lattice is a gapless spin 
liquid for $0.45<J_2/J_1<0.5$ and a (gapped) valence-bond solid for $0.5<J_2/J_1<0.6$.~\cite{Gong-2014} Here, we present 
a variational Monte Carlo method to assess ground-state properties of these models. The starting variational wave functions 
can be substantially improved by the application of few Lanczos steps, which can be easily affordable for rather large 
clusters (i.e., few hundreds sites). Moreover, the calculation of both the energy and its variance makes it possible 
to perform a zero-variance extrapolation to estimate exact energies, not only for the ground state but also for low-energy 
excitations.~\cite{Hu-2013,Iqbal-2013,Iqbal-2014}

\section{Variational wave functions}

The variational wave functions are defined through the mean-field Hamiltonian for the Abrikosov-fermion representation 
of the spin-$1/2$ operators:~\cite{Baskaran-1988}
\begin{equation}\label{eq:meanfield}
{\cal H}_{MF} = \sum_{i,j,\sigma} (t_{i,j} +\mu\delta_{ij}) c^\dag_{i,\sigma} c_{j,\sigma}
+ \sum_{i,j} (\eta_{i,j} +\zeta\delta_{ij}) (c^\dag_{i,\uparrow} c^\dag_{j,\downarrow} + 
c^\dag_{j,\uparrow} c^\dag_{i,\downarrow}) + h.c.,
\end{equation}
where for each bond $(i,j)$ there are hopping $t_{i,j}$ and/or pairing $\eta_{i,j}$ terms; on-site terms, i.e., a 
real chemical potential $\mu$ and/or a complex pairing $\zeta$, may be also considered. Given any eigenstate 
$|\Psi_{MF}\rangle$ of the mean-field Hamiltonian~(\ref{eq:meanfield}), a physical state for the spin model can be 
obtained by a projection onto the subspace with one fermion per site:
\begin{equation}\label{eq:psivar}
|\Psi_v\rangle = {\cal P}_G |\Psi_{MF}\rangle,
\end{equation}
where ${\cal P}_G = \prod_i (n_{i,\uparrow}-n_{i,\downarrow})^2$ is the Gutzwiller projector, 
$n_{i,\sigma}=c^\dag_{i,\sigma} c_{i,\sigma}$ being the local density. The flexibility of this approach is given by the 
fact that {\it different} spin liquids, having for example $U(1)$ or $Z_2$ gauge structure and gapped or gapless spinon 
spectrum, may be obtained by changing the pattern of the $t_{i,j}$'s and the $\eta_{i,j}$'s.~\cite{Wen-2002}
Moreover, dimerized or chiral states can be also easily obtained within this approach.~\cite{Iqbal-2011,Bieri-2014}

In the following, we will consider the square lattice with $J_2/J_1=0.5$ and the Kagome lattice with both $J_2=0$
and $J_2=0.25$, corresponding to cases where the high frustration may stabilize fully symmetric spin liquids.
For the square lattice, we will consider a state that is obtained by taking a real pairing $\eta_{xy}$ (with $d_{xy}$ 
symmetry) on top of the $U(1)$ state with nearest-neighbor hopping $t$ and real pairing $\eta_{x^2-y^2}$ (with
$d_{x^2-y^2}$ symmetry). The $d_{xy}$ term is responsible for breaking the $U(1)$ gauge symmetry down to $Z_2$. 
Restricting this coupling along the $(\pm 2,\pm 2)$ bonds implies four Dirac points in the mean-field 
spectrum.~\cite{Hu-2013} For the Kagome lattice, we will consider the $U(1)$ state that is described by nearest-neighbor 
real hoppings such to have $0$ magnetic flux through triangles and $\pi$ flux through hexagons.~\cite{Ran-2007} 
At the mean-field level, there are gapless excitations at two Dirac points. A slight improvement of the energy can be 
achieved by considering a next-nearest-neighbor real hopping.~\cite{Iqbal-2011} A gapped wave function, corresponding to 
a $Z_2$ spin liquid (labelled by $Z_2[0,\pi]\beta$) can be obtained by adding on-site and next-nearest-neighbor pairing 
terms.~\cite{Lu-2011} A gapless spin liquid with a large Fermi surface (labeled by uniform RVB) is obtained when no 
magnetic fluxes are present.

Within this formalism, it is straightforward to construct not only an Ansatz for the ground state but also for 
low-energy excitations. In this respect, it is useful to consider a particle-hole transformation for the down electrons 
on the mean-field Hamiltonian~(\ref{eq:meanfield}), i.e., $c^\dag_{i,\downarrow} \to c_{i,\downarrow}$, such that the 
transformed Hamiltonian conserves the total number of particles. Then, the ground state is obtained by filling the 
lowest $N$ orbitals, with suitable boundary conditions (either periodic or anti-periodic) in order to have a unique 
mean-field state. Spin excitations can be obtained by creating the appropriate Bogoliubov quasi-particles (spinons) and 
possibly switching boundary condition. Here, we will consider only the case of a $S=2$ excitation with momentum 
$k=(0,0)$, for both the square and the Kagome lattices. By computing separately the energies of the $S=0$ and $S=2$ 
states, the spin gap $\Delta_2$ is studied as a function of the cluster size.

\begin{figure}[t]
\begin{center}
\includegraphics[width=5.5in]{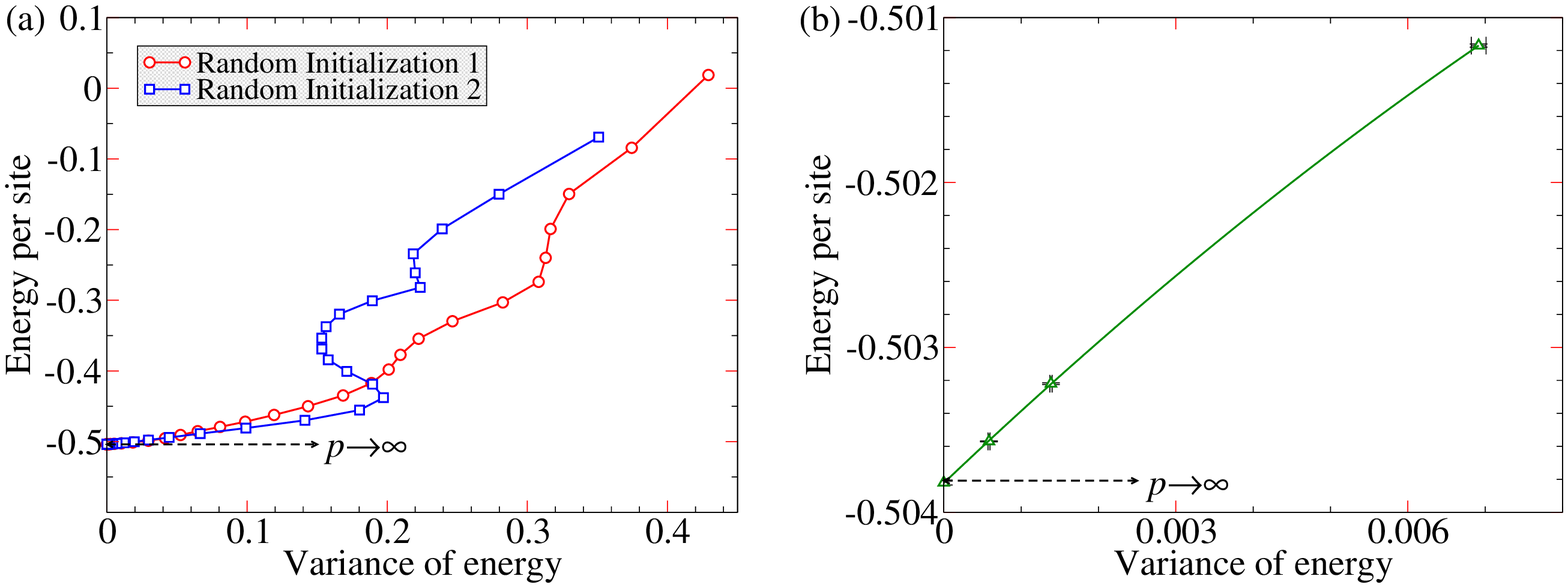}
\end{center}
\caption{\label{fig:square36}
The energy per site versus the energy variance per site for the $J_1{-}J_2$ Heisenberg model on the square lattice with 
$J_2/J_1=0.5$ and $L=6$. Exact Lanczos diagonalizations starting from two random initializations (a) and quantum Monte 
Carlo calculations starting from the best variational state~(\ref{eq:psivar}) (b) are reported.}
\end{figure}

\begin{figure}[t]
\begin{minipage}{18pc}
\includegraphics[width=2.8in]{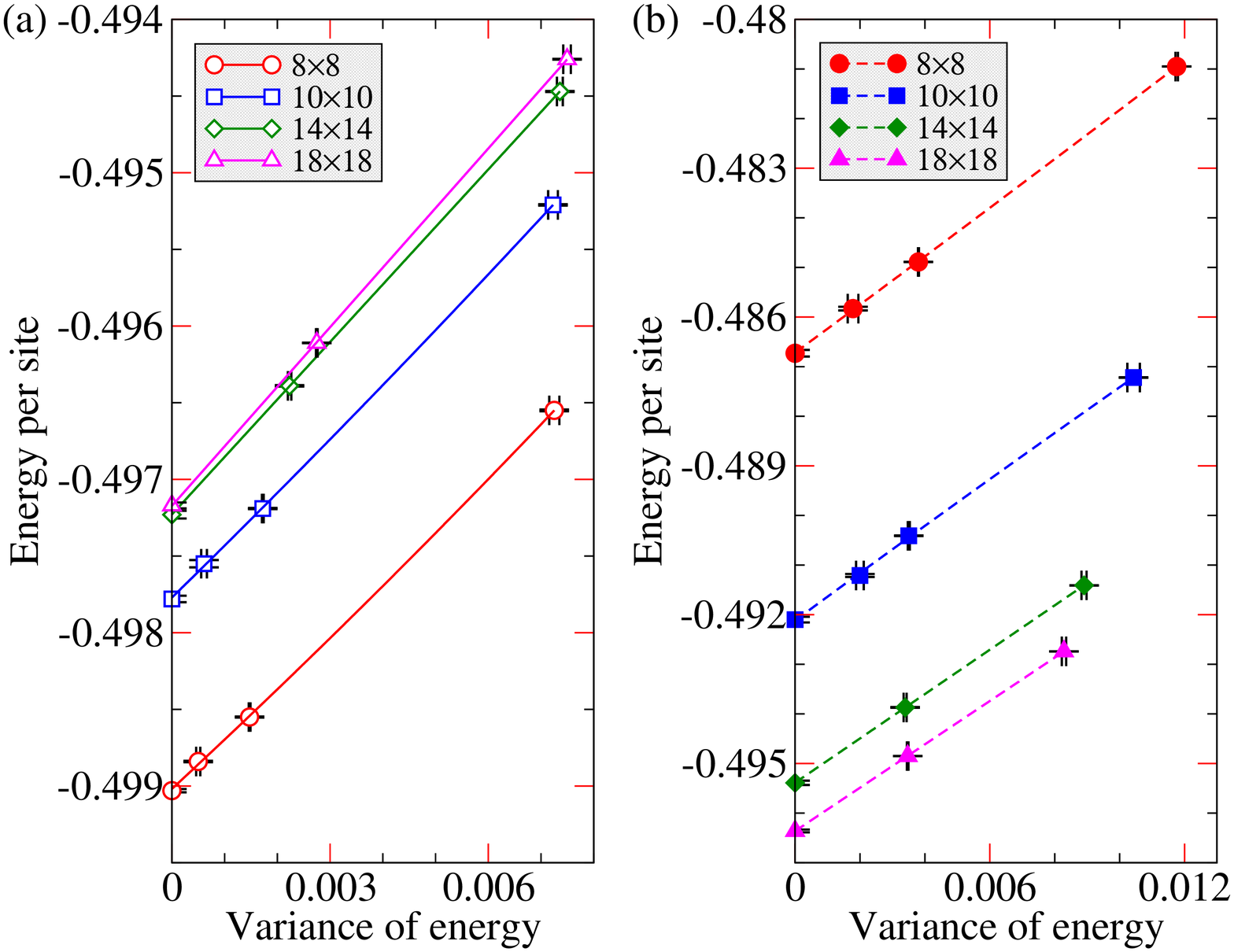}
\caption{\label{fig:square}
Energies per site for $S=0$ (a) and $S=2$ (b) versus the energy variance per site for the Heisenberg model on the square 
lattice with $J_2/J_1=0.5$. The initial state is the best variational Ansatz~(\ref{eq:psivar}). The variance extrapolated 
results are also shown.}
\end{minipage}\hspace{2pc}
\begin{minipage}{18pc}
\includegraphics[width=2.8in]{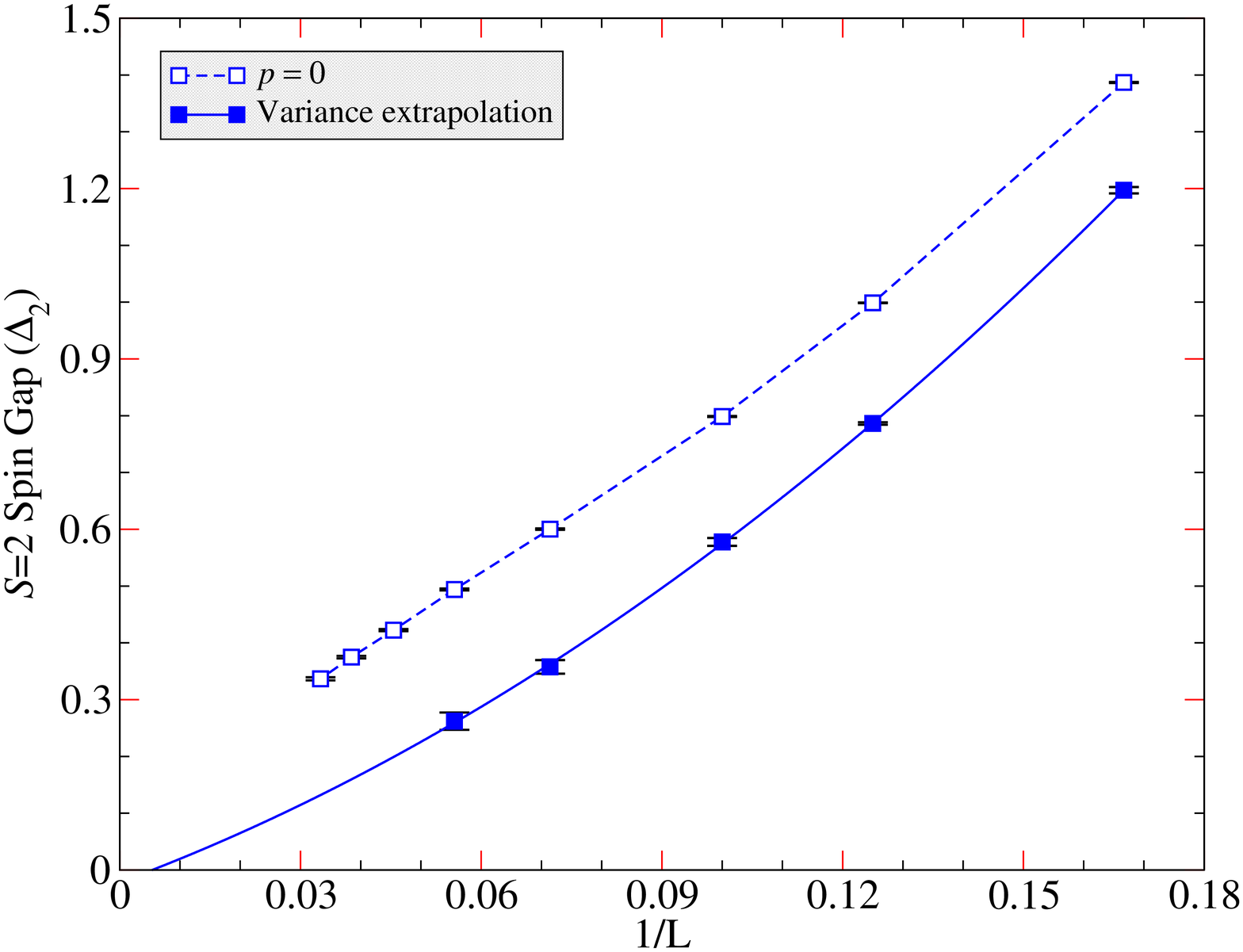}
\caption{\label{fig:squarefinal}
The size scaling of the $S=2$ spin gap for the best variational wave function and the zero-variance extrapolation on the 
square lattice with $J_2/J_1=0.5$. The thermodynamic extrapolation gives $\Delta_2=-0.04(5)$.}
\end{minipage}
\end{figure}

\section{Lanczos step procedure}

In order to systematically improve the variational wave functions, we can apply a number $p$ of Lanczos steps to
the starting variational wave function:
\begin{equation}\label{eq:psilan}
|\Psi_p\rangle = \left ( 1 + \sum_{m=1}^{p} \alpha_m {\cal H}^m \right ) |\Psi_v\rangle,
\end{equation}
where $\alpha_m$ are $p$ additional variational parameters. Clearly, whenever $|\Psi_v\rangle$ is not orthogonal 
to the exact ground state, $|\Psi_p\rangle$ converges to it for large $p$. Unfortunately, on large sizes, only few 
steps can be efficiently afforded: here, we consider the case with $p=1$ and $p=2$ ($p=0$ corresponds to the original 
variational wave function). Furthermore, an estimate of the exact energy may be obtained by the variance extrapolation.
Indeed, for accurate variational states $|\Psi_p\rangle$ with energy $E_p$ and variance $\sigma_p^2$, it is easy to 
prove that
\begin{equation}\label{eq:extrapol}
E_p \approx E_{\rm ex}+{\rm const} \times \sigma_p^2,
\end{equation}
where:
\begin{eqnarray}
E_p &=& \frac{\langle \Psi_p|{\cal H}|\Psi_p\rangle}{N}, \\
\sigma_p^2 &=& \frac{\langle \Psi_p|{\cal H}^2|\Psi_p\rangle-\langle \Psi_p|{\cal H}|\Psi_p\rangle^2}{N},
\end{eqnarray}
are the energy and variance per site, respectively. Therefore, the exact energy $E_{\rm ex}$ may be extracted by fitting 
$E_p$ versus $\sigma_p^2$ and performing a zero-variance extrapolation.

It should be emphasized that the Lanczos step procedure of Eq.~(\ref{eq:psilan}) is not size consistent if $p$ is
not increased with the number of sites $N$. Indeed, both the energy and variance improvements with respect to the
original state $|\Psi_v\rangle$ vanish for $N\to \infty$ and fixed $p$. Nevertheless, it is remarkable that a sizable
improvement is obtained even for rather large clusters with few hundred sites. By contrast, the zero-variance 
extrapolation remains size consistent, as shown below.

\section{Monte Carlo sampling}

By using the variational Monte Carlo technique, the properties of the Gutzwiller-projected wave function can be easily 
assessed. Indeed, expectation values of any operator ${\cal O}$, including ${\cal H}$, are given by:
\begin{equation}\label{eq:average}
\frac{\langle \Psi_p| {\cal O} |\Psi_p\rangle}{\langle \Psi_p|\Psi_p\rangle} = 
\sum_x \frac{\langle \Psi_p|x\rangle \langle x|{\cal O} |\Psi_p\rangle}{\langle \Psi_p|\Psi_p\rangle} =
\sum_x \frac{|\langle \Psi_p|x\rangle|^2}{\langle \Psi_p|\Psi_p\rangle} 
\frac{\langle x|{\cal O} |\Psi_p\rangle}{\langle x|\Psi_p\rangle},
\end{equation}
where $\{|x\rangle\}$ represents a complete and orthogonal basis. By interpreting 
$P(x)=\langle \Psi_p|x\rangle|^2/\langle \Psi_p|\Psi_p\rangle$ as a probability distribution, a Markov chain can 
be constructed to sample Eq.~(\ref{eq:average}). In particular, the Gutzwiller projector is automatically implemented
by taking $\{|x\rangle\}$ written in terms of electron configurations in real space with only singly-occupied sites.
Within the fermionic representation, an efficient Metropolis algorithm can be implemented with local electron moves, 
e.g., spin flips, $|x\rangle \to |x^\prime\rangle$. For $p=0$, whenever few electrons are moved, the computational 
cost of $P(x^\prime)/P(x)$ only requires $O(1)$ operations, while the updating when a new configuration is accepted 
requires $O(N^2)$ operations. In order to have uncorrelated electron configurations, $N$ moves must be done, so that 
the variational algorithm scales like $O(N^3)$. When the first Lanczos step is implemented, the only extra cost is 
that $P(x^\prime)/P(x)$ scales as $O(N)$, while the second Lanczos step gives $O(N^2)$; in both cases the updating 
algorithm remains $O(N^2)$. Therefore, up to $p=2$, the variational calculations have the same computational effort 
as the $p=0$ case. In some cases, in order to have a stable simulation for $p$ Lanczos steps, it is important to adopt 
a regularization scheme to avoid vanishingly small determinants. Here, we consider to sample configurations $|x\rangle$ 
such that: 
\begin{equation}
\sum_{x^\prime \ne x} \left| \frac{\langle x|{\cal H}|x^\prime\rangle \langle x^\prime|\Psi_v\rangle}
{\langle x|\Psi_v\rangle} \right | < \frac{N}{\epsilon}, 
\end{equation}
with $\epsilon$ ranging from $10^{-6}$ to $10^{-5}$.

\section{Results}

We start by showing how the Lanczos step procedure and the variance extrapolation work for a case where exact 
diagonalizations are also available, namely the Heisenberg model on the square lattice with $L=6$ and $J_2/J_1=0.5$. 
In Fig.~\ref{fig:square36}, we report the calculations for the ground state in two different cases, either by starting 
from random initial states (i.e., what is usually done in the Lanczos method) or by initializing with the best 
variational wave function~(\ref{eq:psivar}). While in the former cases, many Lanczos steps are needed to reach a good 
energy per site and, therefore, also the linear regime of Eq.~(\ref{eq:extrapol}), the latter one gives excellent 
results even with two steps, the zero-variance extrapolation being exact within the statistical error. We would like 
to stress that abrupt reductions of the energy, with almost constant variance, or even non-monotonic behaviors, can 
appear when the initial wave function has large overlaps with excited states.

By using the best Gutzwiller projected state, we can extend the Lanczos step procedure to larger systems, where 
exact diagonalizations are not possible. In Fig.~\ref{fig:square}, we report the calculations for both the $S=0$ ground 
state and the $S=2$ excitation. Calculations are performed up to $L=10$ with $p=0$, $1$, and $2$, and up to $L=18$ 
with $p=0$ and $1$. Even though it is visible that the energy/variance gain reduces when increasing $L$, the slope of 
the fit remains similar, allowing a size-consistent zero-variance extrapolation. From these results, the $S=2$ gap is 
shown in Fig.~\ref{fig:squarefinal}. We expect a vanishing gap in the thermodynamic limit for the variational calculations 
with $p=0$, since the $Z_2$ gauge structure is not expected to alter the mean-field properties.~\cite{Wen-2002} 
However, our numerical results up to $L=30$ cannot definitively confirm this fact. Most importantly, the zero-variance
extrapolation for the gap suggests that the frustrated Heisenberg model on the square lattice is gapless for $J_2/J_1=0.5$.
 
\begin{figure}[t]
\begin{center}
\includegraphics[width=5.5in]{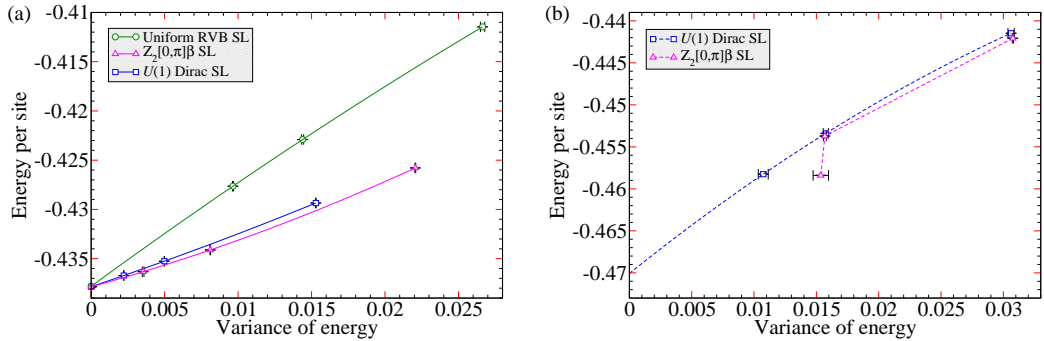}
\end{center}
\caption{\label{fig:kagome48}
Energies per site for the $S=0$ states for the Heisenberg model on the Kagome lattice on $48$ sites (i.e., $L=4$)
for $J_2=0$ (a) and $J_2/J_1=0.25$ (b). Both the $U(1)$ Dirac spin liquid and the gapped $Z_2[0,\pi]\beta$ (for which 
the next-nearest-neighbor pairing $\eta$ is fixed to 1) spin liquid are reported; for $J_2=0$, the uniform RVB state 
with vanishing magnetic fluxes is also reported.}
\end{figure}

\begin{figure}[t]
\begin{minipage}{18pc}
\includegraphics[width=2.7in]{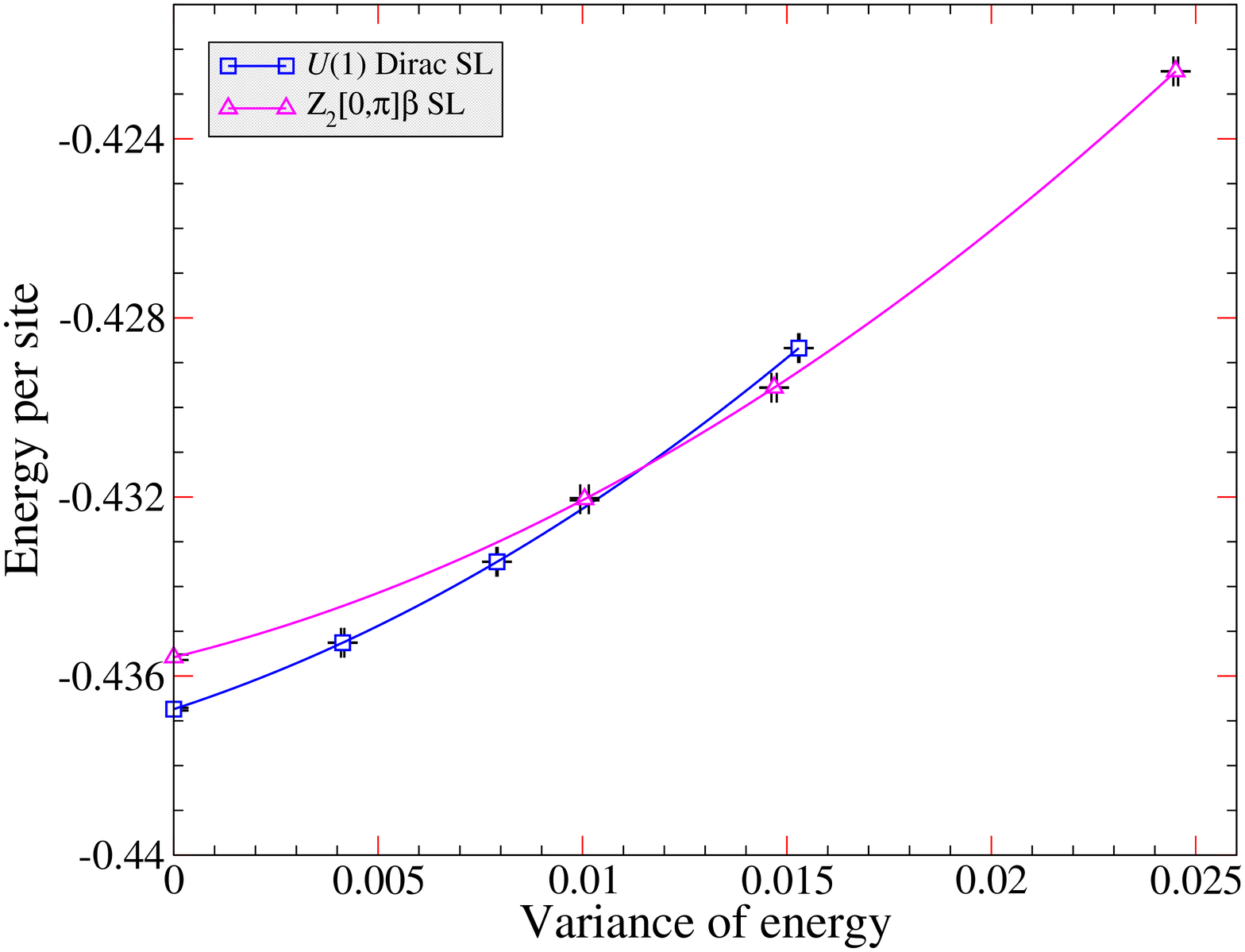}
\caption{\label{fig:kagomej20}
The energy per site versus the variance per site on the Kagome lattice with $L=8$ and $J_2=0$ for the $U(1)$ Dirac 
spin liquid and the gapped $Z_2[0,\pi]\beta$ spin liquid.}
\end{minipage}\hspace{2pc}
\begin{minipage}{18pc}
\includegraphics[width=2.6in]{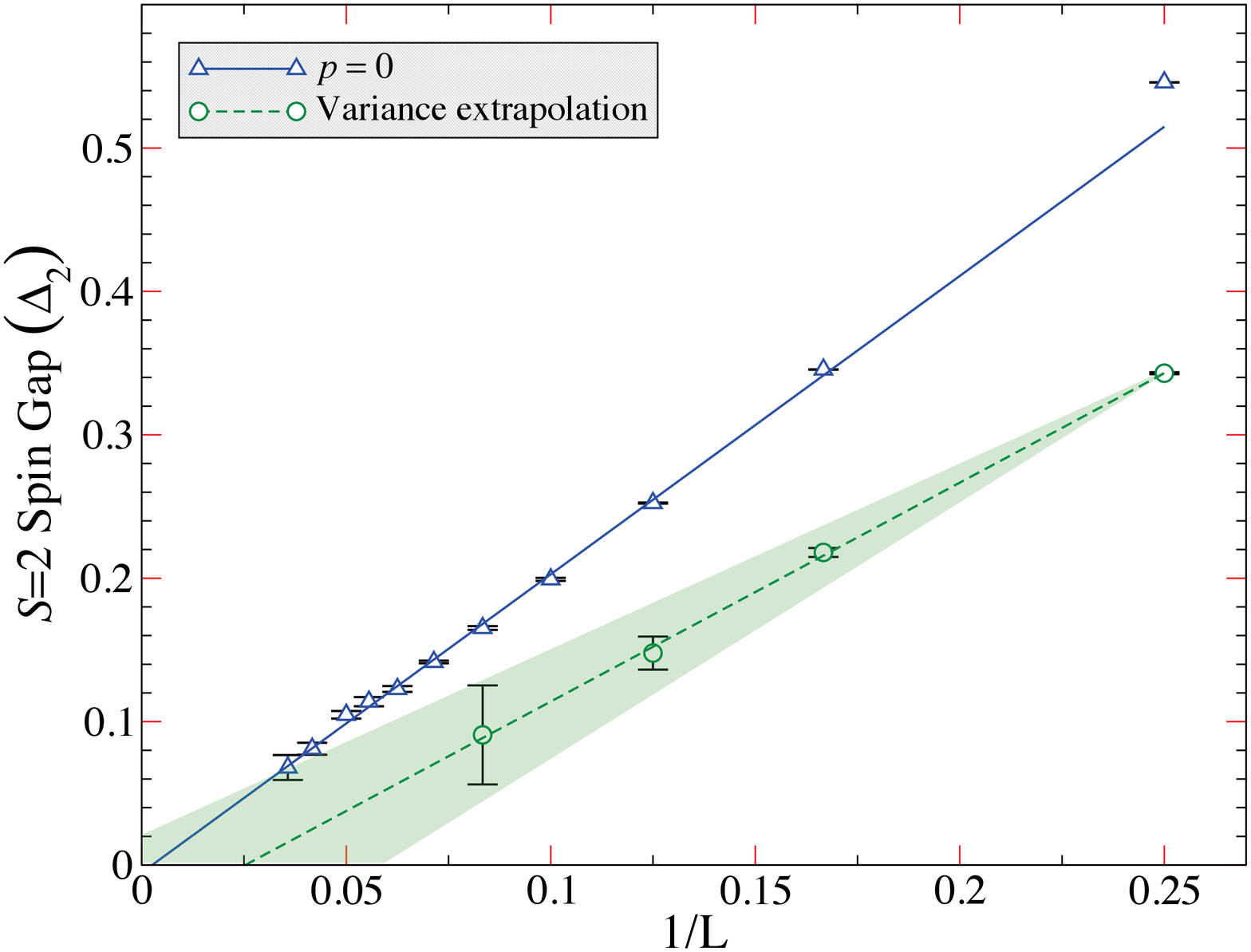}
\caption{\label{fig:kagomefinal}
The size scaling of the $S=2$ spin gap for the $U(1)$ Dirac state and the zero-variance extrapolation on the Kagome 
lattice with $J_2=0$. The thermodynamic extrapolation gives $\Delta_2=-0.04(6)$.}
\end{minipage}
\end{figure}

Let us move to the Kagome lattice. In Fig.~\ref{fig:kagome48}, we report the variance extrapolation for a small
lattice with $48$ sites (i.e., $L=4$) for both $J_2=0$ and $J_2/J_1=0.25$. For the former case, no matter what is
the initial state (gapless with Dirac points or a large Fermi surface, or even gapped), the extrapolated energy is the 
same, within the statistical errors. Notice, however, that the actual values of the initial energies and variances are 
quite different for these three cases. By contrast, for the case with $J_2=0.25$, a rather different behavior is found 
when starting from the $U(1)$ Dirac state or the $Z_2$ gapped one. Remarkably, although the latter one gives a slightly 
better variational energy for $p=0$, the Lanczos extrapolation performs much better for the Dirac state, where a smooth 
fit is possible. Instead, for the $Z_2[0,\pi]\beta$ state, the second Lanczos step makes a sensible energy gain, while 
the variance does not improve. This behavior is reminiscent of what is seen when the initial wave function has some large
overlap with excited states (see Fig.~\ref{fig:square36} for the square lattice). The evidence that the $Z_2[0,\pi]\beta$ 
state is not a good approximation even when $J_2=0$ is reported in Fig.~\ref{fig:kagomej20}. For $L=8$, the Lanczos 
extrapolation of the $U(1)$ Dirac spin liquid gives a better energy than the one obtained from the $Z_2$ state.

By performing a size-scaling extrapolation of the gap by using the $U(1)$ Dirac state, we obtain the results shown in 
Fig.~\ref{fig:kagomefinal}. First of all, it is clear that the Gutzwiller projector does not open a spin gap on top
of the Dirac state: this is a non-trivial outcome, given the fact that a $U(1)$ gauge structure may give rise to 
strong interactions among spinons that may completely change the mean-field picture. Most importantly, also after the 
Lanczos step extrapolation, the system is compatible with a gapless spectrum.

\section{Conclusions}

In summary, we have shown that extremely accurate calculations for the energy per site are possible in frustrated
two-dimensional lattice by using variational wave functions that are built from Abrikosov fermions. Few Lanczos steps 
can be applied to the pure variational state, giving a remarkable energy gain, also for clusters with few hundreds 
sites. Moreover, performing a zero-variance extrapolation, it is possible to obtain accurate values of the exact 
energies, not only for the ground state but also for excited states.

\end{document}